# Charge density wave and weak Kondo effect in a Dirac semimetal CeSbTe

Peng Li[1,2], Baijiang Lv[1], Yuan Fang[1,2], Wei Guo[3], Zhongzheng Wu[1,2], Yi Wu[1,2], Cheng-Maw Cheng[4], Dawei Shen[5], Yuefeng Nie[3,6], Luca Petaccia[7], Chao Cao[8*], Zhu-An Xu[1,6*], Yang Liu[1,2,6*]

[1]*Zhejiang Province Key Laboratory of Quantum Technology and Device, Department of Physics, Zhejiang University, Hangzhou, Mainland China*
[2]*Center for Correlated Matter, Zhejiang University, Hangzhou, Mainland China*
[3]*National Laboratory of Solid State Microstructures, College of Engineering and Applied Sciences, Nanjing University, Nanjing, Mainland China*
[4]*National Synchrotron Radiation Research Center, Hsinchu, Taiwan*
[5]*State Key Laboratory of Functional Materials for Informatics and Center for Excellence in Superconducting Electronics, SIMIT, Chinese Academy of Science, Shanghai, Mainland China*
[6]*Collaborative Innovation Center of Advanced Microstructures, Nanjing University, Nanjing, Mainland China*
[7]*Elettra Sincrotrone Trieste, Strada Statale 14 km 163.5, 34149 Trieste, Italy*
[8]*Department of Physics, Hangzhou Normal University, Hangzhou, Mainland China*
*Corresponding authors:* ccao@hznu.edu.cn, zhuan@zju.edu.cn, yangliuphys@zju.edu.cn

## Abstract

Using angle-resolved photoemission spectroscopy (ARPES) and low-energy electron diffraction (LEED), together with density-functional theory (DFT) calculation, we report the formation of charge density wave (CDW) and its interplay with the Kondo effect and topological states in CeSbTe. The observed Fermi surface (FS) exhibits parallel segments that can be well connected by the observed CDW ordering vector, indicating that the CDW order is driven by the electron-phonon coupling (EPC) as a result of the nested FS. The CDW gap is large (~0.3 eV) and momentum-dependent, which naturally explains the robust CDW order up to high temperatures. The gap opening leads to a reduced density of states (DOS) near the Fermi level ($E_F$), which correspondingly suppresses the many-body Kondo effect, leading to very localized 4$f$ electrons at 20 K and above. The topological Dirac cone at the $X$ point is found to remain gapless inside the CDW phase. Our results provide evidence for the competition between CDW and the Kondo effect in a Kondo lattice system. The robust CDW order in CeSbTe and related compounds provide an opportunity to search for the long-sought-after axionic insulator.

**Introduction**

The charge density wave (CDW), a periodic modulation of the electronic charge density in a crystal, has been studied for over half a century and remains an important subject in condensed matter physics. The CDW order can compete with other ordered phases, including superconductivity and magnetism [1,2,3,4], leading to a variety of interesting phenomena. Conventionally, CDW orderings in low-dimensional systems can be explained by electron-phonon coupling (EPC) due to the FS nesting, where a CDW ordering vector connects the parallel portions of the FS and leads to gap opening with reduced energy. Since the nested FS is better fulfilled in lower dimensions, the CDW ordering normally occurs in quasi one-dimensional (1D) or two-dimensional (2D) systems. However, in some other CDW systems where the condition of the FS nesting is not well satisfied, e.g., in some transition metal dichalcogenides, other CDW mechanisms have been proposed, such as excitonic interaction [5,6], Jahn-Teller effect [7,8], momentum-dependent EPC [9], and strong electron correlation [10], which are still hotly debated. CDW can also occur in strongly correlated Kondo lattice systems [11,12], but how the CDW ordering interacts with the many-body Kondo effect remains largely unexplored. Theoretically, the many-body Kondo screening [13,14] requires mobile conduction electrons near $E_F$, while the CDW order normally reduces the DOS near $E_F$ via gap opening. A natural question would be: will the Kondo effect be largely suppressed by the CDW ordering? The quasiparticle dispersion and FS obtained from ARPES can be essential to addressing this question.

On the other hand, CDW orderings in topological semimetals have also attracted considerable interest recently [15,16]. Topological semimetals are special types of semimetals, exhibiting low-energy excitations that are analogous to high-energy elementary particles [17]. They can be classified into Dirac semimetals [18,19], Weyl semimetals [20,21,22,23,24], nodal-line semimetals [25] and semimetals with exotic degeneracies [26,27]. Dirac semimetals host fourfold degenerate Dirac points with linear dispersion along three momentum directions, which are protected by crystal symmetries. Upon breaking time reversal or inversion symmetry, the Dirac point will split into a pair of Weyl points with opposite chiralities, resulting in Weyl semimetals. It was theoretically proposed that an axionic insulator could be induced from a Weyl semimetal in the presence of the CDW ordering, where the CDW ordering gapped out the Weyl points with different chiralities, leading to axionic excitations within the gap [28,29,30]. Similar axionic phases have also been predicted in Dirac semimetals [31]. In these cases, the phase of the CDW order parameter could be coupled to the electromagnetic field in an analogous way to axion in quantum field theory, giving rise to unusual magnetoelectric effects. Due to the stringent requirement of both the CDW ordering and the Dirac/Weyl band-crossing, the experimental evidence for the axionic phase is very scarce. Only very recently, was such experimental evidence reported in the CDW-ordered Weyl semimetal $(TaSe_4)_2I$ from magneto-conductance measurements [32], providing fascinating opportunities to study this novel quasiparticle. Nevertheless, band structure measurements are urgently needed to understand the underlying physics and the interplay between CDW and topology.

In this paper we report the observation of CDW ordering and its interplay with the Kondo effect in a Dirac semimetal CeSbTe, which was proposed to host different Dirac/Weyl states due to its nonsymmorphic crystal symmetry and tunable magnetic order [33]. The tetragonal phase CeSbTe is isostructural to ZrSiS, a famous nodal-line semimetal protected by crystal symmetry and negligible spin orbit coupling (SOC) [34,35]. Since SOC can no longer be ignored in CeSbTe, it is actually a

Dirac semimetal in the paramagnetic phase because the Dirac point at the $X$ point is protected by the nonsymmorphic crystal symmetry [33]. CeSbTe also exhibits an antiferromagnetic (AFM) transition at $T_N \sim 2.7$ K, and a metamagnetic transition to the ferromagnetic (FM) phase under a small magnetic field $\sim 0.3$ T. These magnetic transitions break time-reversal symmetry and can induce Weyl points or band crossings with even higher degeneracies [33]. Specific heat measurements at low temperatures indicate a moderately enhanced Sommerfeld coefficient $\gamma \sim 41$ mJ mol$^{-1}$·K$^{-2}$, with an estimated Kondo temperature $T_k \sim 1.3$ K [36]. This implies active Kondo screening, although the Kondo effect is probably suppressed due to its low carrier density [36]. As we shall show below, the presence of CDW provides an interesting opportunity to study the interplay between CDW, Kondo effect and nontrivial band topology in CeSbTe and to search for new quantum phases arising from their interactions.

**Experimental and computational details**

The single crystals used in this study were synthesized by the vapor transport technique and reported in a previous paper [36]. ARPES measurements of single crystals of CeSbTe(001) (with a typical lateral size of 1 × 1 mm$^2$) were performed at low temperatures (25-40 K), after being cleaved *in situ* with a base pressure < 1 × 10$^{-10}$ Torr. The low energy ultraviolet ARPES measurements (< 40 eV) were carried out at the BaDE1Ph beamline [37] of the Elettra Synchrotron Light Source, and measurements with higher photon energies were performed at the BL03U beamline of Shanghai Synchrotron Radiation Facility and the BL 21B1 beamline at Taiwan Light Source. The typical energy resolution of the ARPES measurements is about 20 meV, and the momentum resolution is ~0.02 Å$^{-1}$. LEED measurements were performed with an incident electron energy of 100 eV. All data presented in this paper were taken within a few hours after cleavage, to ensure our results were not affected by sample aging.

Electronic structure calculations in this work were performed using DFT and a plane-wave basis projected augmented wave method, as implemented in the Vienna *ab initio* simulation package (VASP) [38]. The Perdew, Burke and Ernzerhof (PBE) generalized gradient approximation for exchange correlation potential was used for the DFT calculation. The 4$f$ electrons were treated as localized core electrons and SOC was taken into account in all DFT calculations. An energy cutoff of 300 eV and a 12 × 12 × 6 gamma-centered k-mesh were employed to converge the calculation to 1 meV/atom.

**Results and discussion**

The tetragonal CeSbTe crystallizes in the nonsymmorphic space group P4/nmm (No. 129) as shown in Fig. 1(a), where the Ce-Te bilayers are sandwiched between the layers of Sb forming square nets. Its bulk Brillouin zone (BZ) and projected surface BZ are shown in Fig. 1(b): the bulk high symmetry points $\Gamma$, M, X ($k_z = 0$) and Z, A, R ($k_z = \pi$) are projected onto the $\bar{\Gamma}$, $\bar{M}$ and $\bar{X}$ points in surface BZ. The sample quality is checked by the momentum-integrated energy distribution curve (EDC) shown in Fig. 1(c), which reveals expected core levels including Te 4$d$, Sb 4$d$, Ce 5$s$, 5$p$ and 4$f$ peaks. The resistivity shows an overall insulating behavior, with a broad hump at ~75 K, which has been explained by the crystal electric field excitation in a previous report [36]. An additional upturn in the resistivity can be observed below 10 K, which can be attributed to the Kondo scattering ($T_k \sim 1.3$

K). Upon further lowering the temperature, a small kink occurs at 2.62 K, corresponding to the AFM transition [33,36]. We note that the weak insulating behavior starting from the room temperature (RT) cannot be easily explained by the normal-state band structures from DFT. It is more likely caused by the gap opening by CDW, as we shall discuss below.

Now we discuss its electronic structure from ARPES measurements. We note that ARPES results have been already reported on this compound [33,39], emphasizing the topological aspect of the band structure. Here we focus on the fermiology and its connection with the CDW order. The FS map is shown in Fig. 2(a), with the corresponding high symmetry cuts shown in Fig. 2(b,c). The band structures from DFT calculations are overlaid on top of the ARPES spectra in Fig. 2(b,c), showing good agreement. Note that the DFT calculations assume fully localized 4$f$ electrons, which implies that the Ce 4$f$ electrons are localized and do not participate in the FS, as we shall discuss in further detail below. There are several bands in the ARPES data that cannot be explained by the bulk DFT calculations (labelled SS, SS1 and SS2), which can be attributed to surface states as confirmed by previous slab calculations [39]. Zooming in on the band dispersion near $E_F$ along $\overline{M}-\overline{\Gamma}-\overline{M}$ (see Fig. 2(d)), we can identify two sets of bands, whose intensity diminishes upon getting close to $E_F$ (due to CDW gap opening). The photon energy dependence of the momentum distribution curves (MDCs) at $E_F$ is presented in Fig. 2(e), which exhibits no noticeable $k_z$ dispersion, indicating the quasi-2D nature of the observed bands. This is further supported by the 3D FS plot from DFT calculation shown in Fig. 2(f), which reveals two 2D rhombus-shaped sheets extending along $k_z$ and a small enclosed pocket near $\Gamma$. The inner rhombus sheet (closer to $\Gamma$, referred to as the inner sheet thereafter) is a 2D hole pocket centered at $\Gamma$, while the outer rhombus sheet (referred to as the outer sheet thereafter) actually consists of a large 2D hole pocket centered at $\overline{M}$ and a small quasi-2D electron pocket centered at $\overline{X}$ (see also Fig. 3(c)). These FS sheets are in excellent agreement with the experimental FS map in Fig. 2(a). The small pocket at $\overline{\Gamma}$ predicted by DFT is rather weak experimentally and hence cannot be clearly identified (see also Fig. 3(c)), likely due to its vanishing photoemission matrix element (under $p$-polarized photons).

The inner and outer sheets show largely parallel segments that can be connected by one (symmetry-equivalent) wave vector: this well-nested FS can drive CDW ordering via strong EPC. In fact, the FS in Fig. 2(a) exhibits striking similarity with the FSs from rare earth tri-tellurides (RETe$_3$), such as CeTe$_3$ [40], where CDW orderings with high ordering temperatures have been reported as a result of FS nesting [40,41,42]. This is perhaps not too surprising: CeSbTe shares the same crystal structure as CeTe$_2$ [33,36,43], except that one Te layer is replaced by a Sb layer (resulting in hole doping), while CeTe$_3$ can be viewed electronically as hole-doped CeTe$_2$ due to the extra Te layer [40]. It is interesting to note that many rare earth ditellurides (RETe$_2$) also exhibit CDW orderings driven by FS nesting [43,44,45,46], despite different FS topologies due to chemical potential shifts.

The CDW order was confirmed by LEED experiments as shown in Fig. 3(a), where one can clearly see CDW superstructure peaks in addition to main lattice peaks. The CDW peaks persist from 12 K up to 338 K (the highest temperature we could reach). For quantitative analysis, we plot the diffraction intensity along the [100] direction (the black dashed line in Fig. 3(a)) and the extracted ratio between the reciprocal lattice vector and the CDW ordering vector in Fig. 3(b). The analysis yields a ratio close to 5, with no obvious temperature dependence, implying a nearly commensurate 5×1 CDW order (with two domains rotated 90° with respect to each other) within the probed temperature range. Fig. 3(c) shows the extended FS map, with the calculated 2D FS ($k_z = 0$) overlaid on top. The observed CDW ordering vector, q$_{CDW}$ (white arrows), can connect parallel segments between the inner and outer sheets very well. Since both FS sheets are also 2D, this well-nested FS can naturally explain the strong CDW order in this system.

To obtain the CDW gap size and its momentum dependence, we measured a high-quality FS map near the $\overline{X}$ point, as shown in Fig. 4(a). We extracted the CDW gap along both the inner (D-C) and outer sheets (B-A), by performing a standard symmetrization procedure for gap analysis, where the original EDC is added to its mirror with respect to $E_F$, effectively removing the Fermi-Dirac function. The symmetrized EDCs are summarized in Fig. 4(b), where a clear suppression of DOS can be seen near $E_F$, signaling CDW gap opening. We note that CDW does not completely gap out the DOS at $E_F$, and the leading edge of the gap is rather broad, similar to other $RETe_2$ and $RETe_3$ CDW materials [41,44]. As usual, we define the midpoint position of the leading edge as half of the CDW gap ($\Delta/2$). The evolution of the CDW gap is summarized in Fig. 4(c), which clearly shows an approximately constant gap size except near the $\overline{X}$ point, where the gap suddenly decreases (but remains finite). The reduction of the CDW gap near the $\overline{X}$ point can be explained by the imperfect nesting near this momentum region, which has also been observed in other $RETe_3$ CDW materials [41,42]. The large CDW gap observed in CeSbTe is comparable to $CeTe_2$ (with a half gap ~0.2 eV) [47,48] and $CeTe_3$ (with a half gap ~0.2 eV) [40,49], indicating that their CDW orders share much in common. The large CDW gap also provides a natural explanation for the robustness of the CDW order even at high temperatures.

Having established the CDW order, it is interesting to understand the possible interplay between CDW and the Kondo effect (from the $4f$ electrons in Ce). A previous report showed that the Kondo scattering was present in the system, although the Kondo temperature was found to be low (~1.3 K), which was attributed to the small carrier density as obtained from Hall measurements [36]. As the many-body Kondo screening requires mobile conduction electrons, it is natural to expect that the depleted conduction electrons can suppress the Kondo effect substantially. However, the carrier density deduced from the Hall measurement [36], ~$1.23 \times 10^{21}$ cm$^{-3}$, is one order of magnitude smaller from the value estimated from the Luttinger volume of the two dominant hole-type FS sheets (~$1 \times 10^{22}$ cm$^{-3}$). This indicates that the CDW order indeed reduces the DOS near $E_F$ significantly by opening large (anisotropic) energy gaps. The spectroscopic signature of the Kondo effect is the characteristic Kondo resonance peaks near $E_F$ in the $4f$ spectral function [50,51]. To probe the $4f$-specific spectral function, we performed both off-resonant and on-resonant ARPES spectra near the Ce $M$ edge ($4d \rightarrow 4f$), as shown in Fig. 5. While the off-resonant ARPES spectra (110 eV) were dominated by emission from conduction bands, the $4f$ spectral function was largely enhanced at the resonance condition (122 eV). The on-resonant spectrum featured two broad $4f$ peaks, one at -3.2 eV and the other at -0.7 eV. The deep energy peak (-3.2 eV) can be attributed to the localized $4f$ peak ($4f^0$), whose energy is deeper than most Kondo systems (typically between -3 eV and -2 eV), including prototypical low carrier density Kondo systems such as CeSb and CeBi [52,53]. Based on the single impurity Anderson model, a deeper $4f^0$ peak reflects the more localized nature of $4f$ electrons [51]. Indeed, no Kondo resonance peaks can be observed near $E_F$, indicating that the many-body Kondo process was largely suppressed at the measurement temperature (25 K). The $4f$ peak at -0.7 eV, also observed in $CeTe_2$ and other related Ce-based compounds [47,54,55], is likely due to the final state effect in the photoemission process, where the $4f$ hole after photoexcitation can be filled by a higher-energy conduction electron due to hybridization effect, leading to an energy shift of the outgoing $4f$ electron. Nevertheless, future studies are still needed to confirm its origin.

According to a previous study, the bulk band structure (without considering CDW) features a

Dirac cone near the $X$ point, protected by the nonsymmorphic crystal symmetry [33], i.e., the system is a Dirac semimetal. Similar Dirac cones have also been reported in isoelectronic GdSbTe [56] and NbSbTe [57]. With the observation of the CDW order, it is important to investigate how the CDW order affects the Dirac cone. Figs. 6(a) and (b) illustrate the constant energy contours at several energies and the corresponding curvature maps, respectively. The curvature maps were obtained from the direction-averaged second derivatives of the original maps [58]. The Dirac point is located at -0.37 eV, which can be seen as a gapless band crossing in the $E$= -0.37 eV map in Fig. 6(a,b). The Dirac point can be better visualized in the band dispersion and its second derivative along the $\overline{\Gamma}-\overline{X}-\overline{\Gamma}$ direction (Fig. 6(c,d)), where a characteristic Dirac cone can be clearly seen. The Dirac cone is also in good agreement with the band structure calculation (red dashed curves), i.e., in the normal state without CDW. In particular, we found no evidence of gap opening near the Dirac point. Therefore, it appears that the Dirac cone remains gapless inside the CDW phase, implying that the structural distortion associated with the CDW order likely preserves the nonsymmorphic crystal symmetry (or the distortion-induced gap is too small for experimental detection).

Taken together, here we observe the coexistence of a CDW energy gap near $E_F$ (less than 0.1 eV close to the $X$ point, see Fig. 4(c)) and a gapless Dirac cone (at -0.37 eV) at the $X$ point. If the compound can be further doped (with holes) to shift the Dirac point close to $E_F$, the CDW could gap the Dirac point, making it a possible candidate to search for the long-sought axionic insulator. We infer that the CDW might be a robust ground state in this family of materials, including RETe$_2$, RETe$_3$ and possibly RESbTe, thus enabling chemical tuning.

## Conclusion

By combining ARPES and LEED measurements, together with DFT calculations, we have observed a robust CDW order and weak Kondo effect in a Dirac semimetal CeSbTe, which is driven by EPC from the well-nested 2D FS. The observed CDW ordering vector from LEED, approximately 1/5 of the reciprocal lattice vector, agrees very well with the nesting vector of the experimental FS from ARPES. The measured CDW gap is large and up to ~0.3 eV, thus explaining the robust CDW order up to very high temperatures. The gap is roughly isotropic in momentum space, except near the $X$ point where the imperfect FS nesting leads to a smaller energy gap (less than 0.1 eV). The reduced DOS near $E_F$ as a result of the CDW gap leads to a low carrier density measured from the Hall effect, which is much smaller than the expected value without CDW. The reduced DOS in turn suppresses the many-body Kondo screening at low temperatures, evidenced by the absence of Kondo resonance peaks near $E_F$. The Dirac cone at the $X$ point seems to remain gapless inside the CDW phase, indicating that the nonsymmorphic crystal symmetry is likely preserved within the CDW phase. Considering the robustness of the CDW order and large chemical tunability observed, this class of materials, including RETe$_2$, RETe$_3$, and RESbTe, may serve as an interesting material system to realize the long-sought axionic phase.

*This work is supported by the National Key R&D Program of the MOST of China (Grant No. 2016YFA0300203, 2017YFA0303100), the National Science Foundation of China (No. 11674280, 11774305) and the Science Challenge Program of China. Part of this research used Beam line 03U of the Shanghai Synchrotron Radiation Facility, which is supported by ME2 project under Contract No. 11227902 from National Natural Science Foundation of China. We also acknowledge Elettra*

*Sincrotrone Trieste for providing access to its synchrotron radiation facilities. Finally, we would like to thank Dr. Pei-Yu Chuang, Dr. Zhengtai Liu, Dr. S. Gonzalez and Dr. G. Di Santo for help in the synchrotron ARPES measurements.*

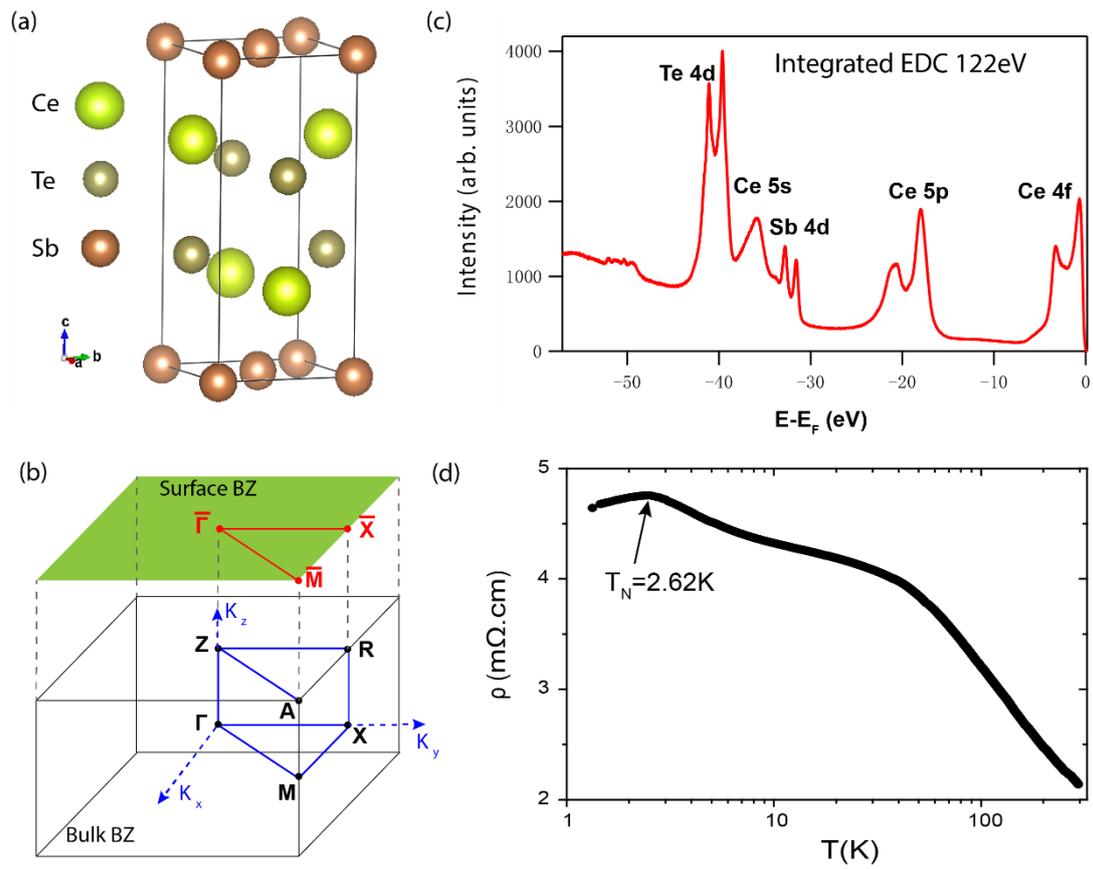

**Figure 1** (a) Crystal structure of CeSbTe. (b) The corresponding bulk and surface BZ. (c) Momentum integrated EDC of CeSbTe taken with 122 eV photons. (d) The resistivity vs temperature in the logarithmic scale. The black arrow indicates the AFM transition at 2.62 K.

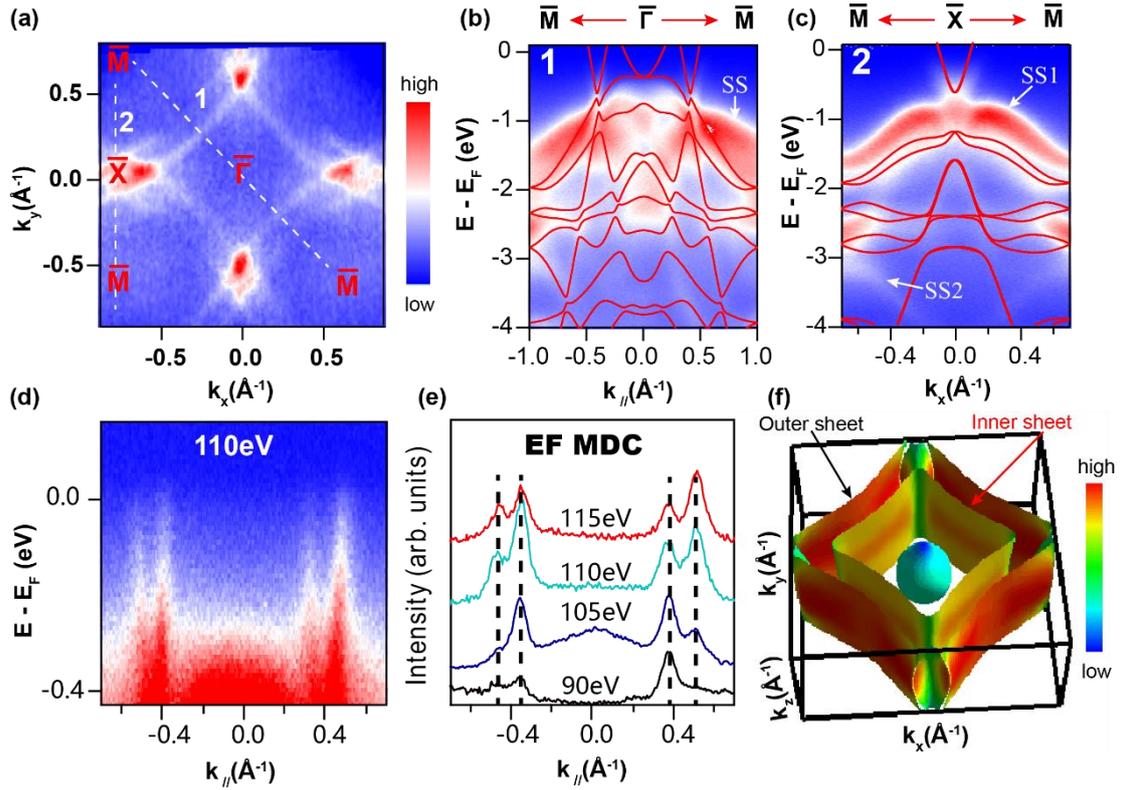

**Figure 2** (a) Experimental FS map at 25 K using 130 eV photons. Two high symmetry cuts corresponding to (b,c) are indicated by white dashed lines. (b,c) ARPES spectra taken along the $\overline{M}-\overline{\Gamma}-\overline{M}$ (b) and $\overline{M}-\overline{X}-\overline{M}$ (c) directions. The bands from DFT calculations (red curves) are overlaid on top of the experimental data. The white arrows indicate surface states (SS, SS1 and SS2) as reported before [39]. The ARPES data was taken at 30 K with 110 eV photons. (d) A zoom-in view of the band dispersion along $\overline{M}-\overline{\Gamma}-\overline{M}$ near $E_F$. (e) MDCs at $E_F$ along $\overline{M}-\overline{\Gamma}-\overline{M}$ under different photon energies. (f) Calculated three-dimensional FS. The inner and outer sheets were labelled by arrows. The color indicates the band velocity.

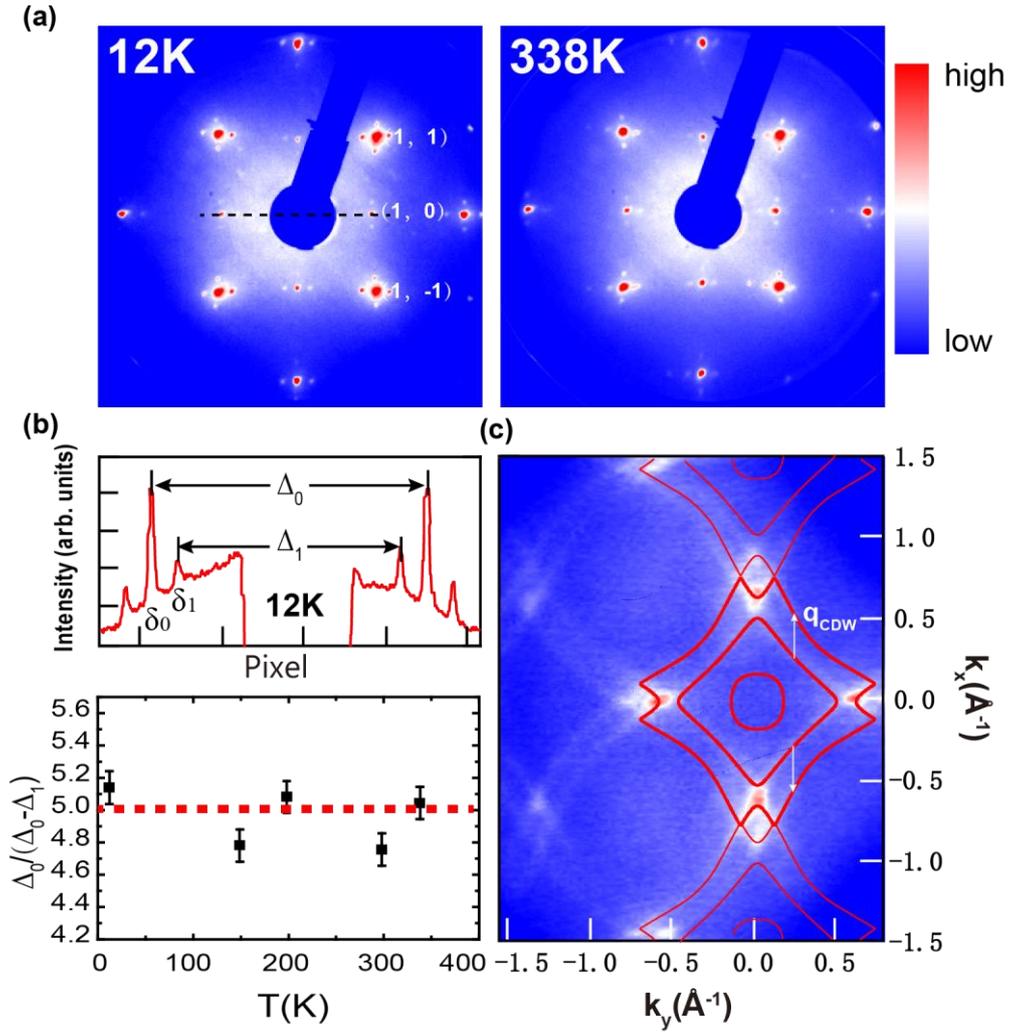

**Figure 3** (a) LEED patterns taken at 12 K and 338 K with an incident electron energy of 100 eV. Three lattice diffraction peaks were labelled. (b) Upper panel: Intensity plot along the horizontal black dashed line in (a), where the lattice peak ($\delta_0$) and the CDW peak ($\delta_1$) are labelled. Bottom panel: temperature dependence of $\Delta_0/(\Delta_0 - \Delta_1)$, where $\Delta_0$ is the distance between the (-1,0) and (1,0) lattice peaks, and $\Delta_1$ is the distance between two CDW peaks. Its value is close to 5 (the red dashed line) within the experimental uncertainty. (c) Extended FS map with calculated 2D FS ($k_z = 0$) overlaid on top (the thick red lines highlight bands within the first BZ). The FS is well nested by the observed CDW ordering vector $q_{CDW}$ (white arrows).

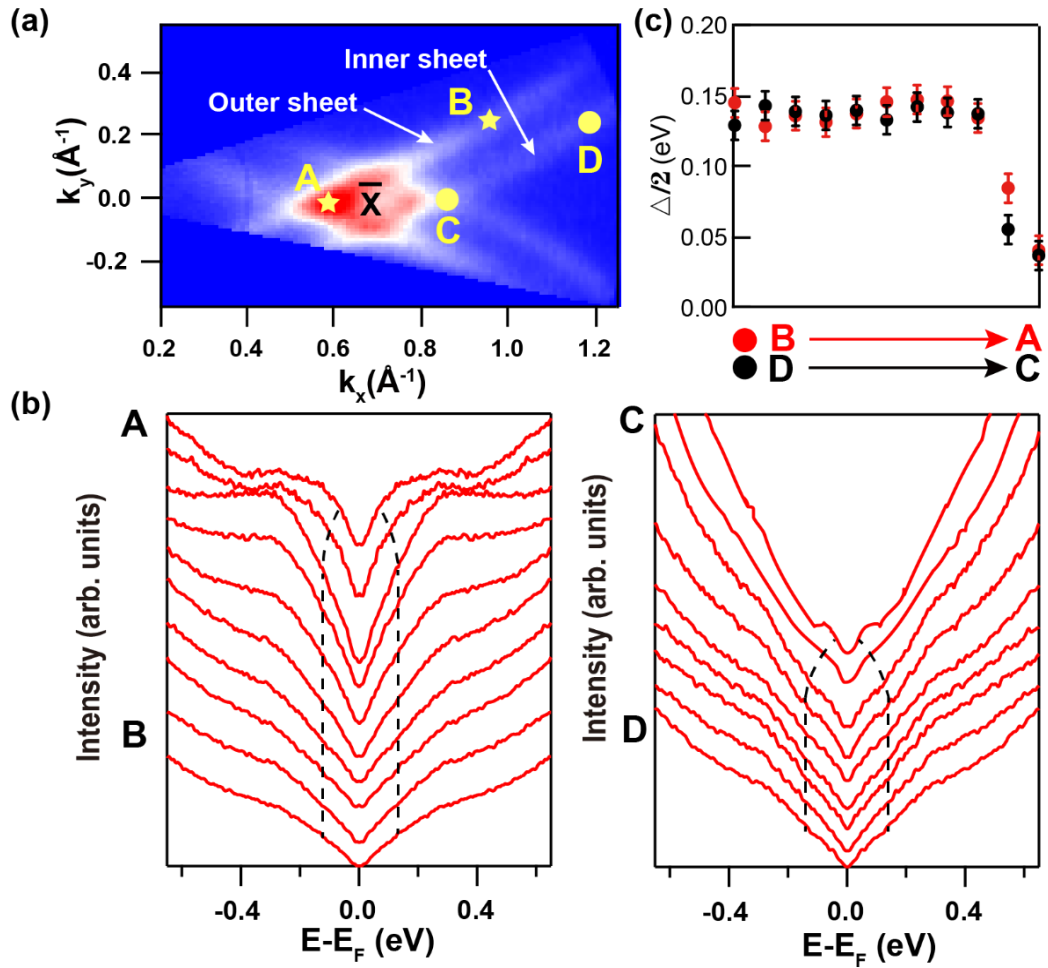

**Figure 4** (a) High resolution FS map taken at 40 K using 28 eV photons. Representative momentum points are labelled, including A and B points (yellow filled stars) along the outer sheet, C and D points (yellow filled circles) along the inner sheet. (b) Symmetrized EDCs along the outer sheet (left) and the inner sheet (right). The black dashed lines indicate the evolution of the midpoint of the leading edge. (c) Extracted one half of the CDW gap along the inner and outer sheets with error bars.

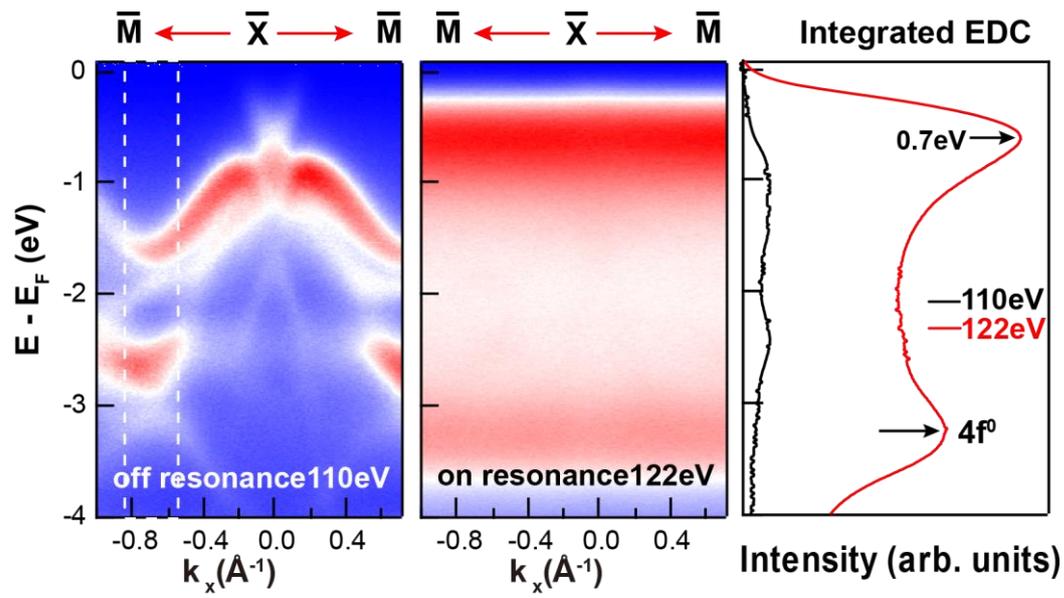

**Figure 5** Off-resonant (110 eV) and on-resonant (122 eV) ARPES spectra along $\overline{M}-\overline{X}-\overline{M}$ at 25 K and corresponding integrated EDCs within the white dashed rectangle (shown in the right panel).

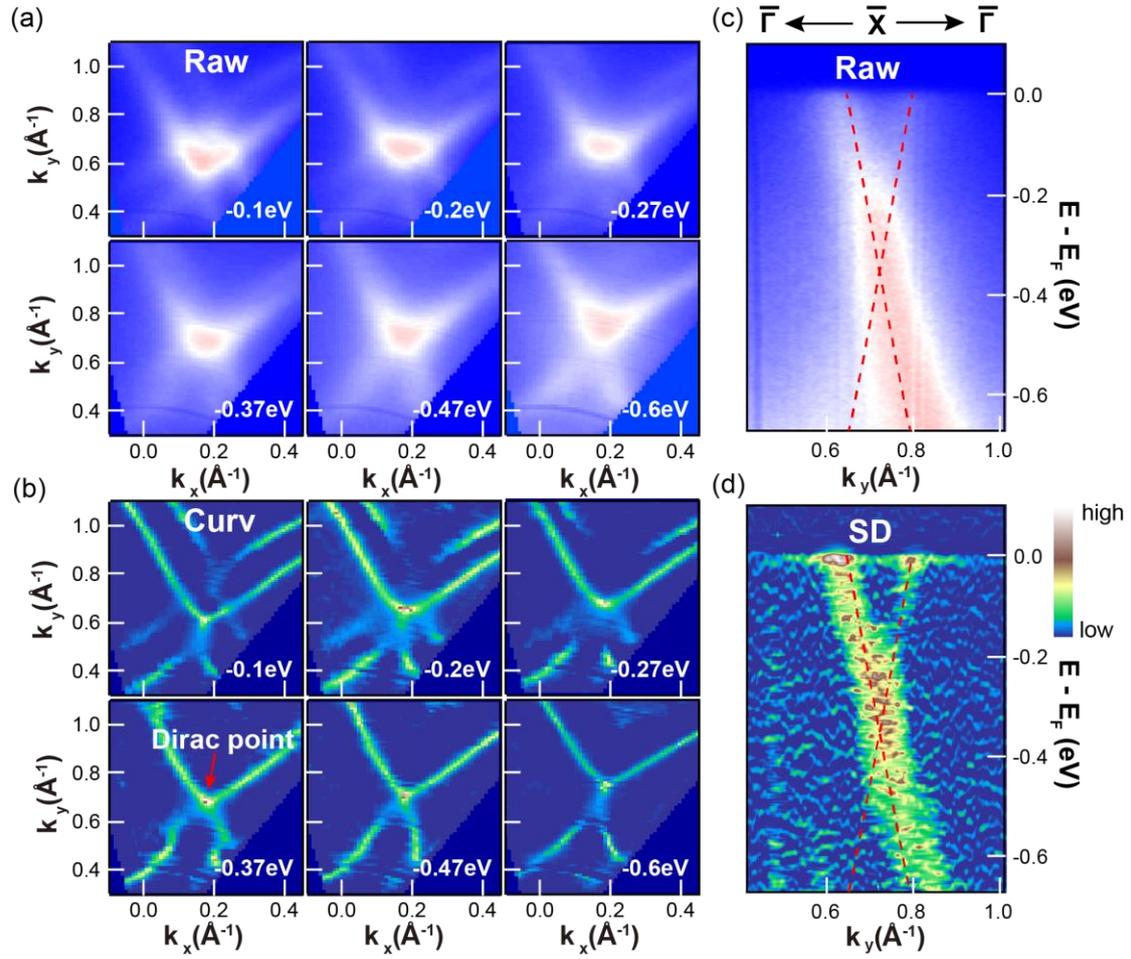

**Figure 6** (a,b) Constant energy maps (a) at several representative energies and their corresponding curvature maps (b), obtained from second derivatives. The Dirac point is labelled by a red arrow. (c,d) ARPES band dispersion (c) and its second derivative (d) along $\bar{\Gamma}-\bar{X}-\bar{\Gamma}$ direction with the calculated bulk Dirac cone overlaid on top (red dashed curves).